\documentclass[twocolumn,amsmath,superscriptaddress]{revtex4}

\usepackage{graphicx}
\usepackage{epsfig}
\begin{document}
\title{Simultaneous observation of free and defect-bound excitons in CH$_3$NH$_3$PbI$_3$ using four-wave mixing spectroscopy}

\author{Samuel A. March}
\affiliation{Department of Physics and Atmospheric Science,
Dalhousie University, Halifax, Nova Scotia B3H 4R2 Canada}

\author{Charlotte Clegg}
\affiliation{Department of Physics and Atmospheric Science,
Dalhousie University, Halifax, Nova Scotia B3H 4R2 Canada}

\author{Drew B. Riley}
\affiliation{Department of Physics and Atmospheric Science,
Dalhousie University, Halifax, Nova Scotia B3H 4R2 Canada}

\author{Daniel Webber}
\affiliation{Department of Physics and Atmospheric Science,
Dalhousie University, Halifax, Nova Scotia B3H 4R2 Canada}

\author{Ian G. Hill}
\affiliation{Department of Physics and Atmospheric Science,
Dalhousie University, Halifax, Nova Scotia B3H 4R2 Canada}

\author{Kimberley C. Hall}
\affiliation{Department of Physics and Atmospheric Science,
Dalhousie University, Halifax, Nova Scotia B3H 4R2 Canada}

\begin{abstract}
{\bf Solar cells incorporating organic-inorganic perovskite, which may be fabricated using low-cost solution-based processing, have witnessed a dramatic rise in efficiencies yet their fundamental photophysical properties are not well understood.  The exciton binding energy, central to the charge collection process, has been the subject of considerable controversy due to subtleties in extracting it from conventional linear spectroscopy techniques due to strong broadening tied to disorder.  Here we report the simultaneous observation of free and defect-bound excitons in CH$_3$NH$_3$PbI$_3$ films using four-wave mixing (FWM) spectroscopy.  Due to the high sensitivity of FWM to excitons, tied to their longer coherence decay times than unbound electron-hole pairs, we show that the exciton resonance energies can be directly observed from the nonlinear optical spectra.  Our results indicate low-temperature binding energies of 13~meV (29~meV) for the free (defect-bound) exciton, with the 16~meV localization energy for excitons attributed to binding to point defects.  Our findings shed light on the wide range of binding energies (2-55~meV) reported in recent years.}
\end{abstract}


\maketitle
The efficiency of solar cells using organometal halide perovskite semiconductor absorber layers has skyrocketed during the past few years, now reaching a level competitive with existing solar cell technologies \cite{NREL:web} while offering a low-cost solution-processed device platform.  This has stimulated intense research into the fundamental photophysical properties of these materials \cite{Wu:2014,Savenije:2014,Sun:2014,He:2016,Wehrenfennig:2014,Tanaka:2003,Hirasawa:1994,Miyata:2015,Galkowski:2016,Lin:2014,SabaNatComm:2014,Even:2014,DInnocenzo:2014,YamadaIEEE:2015,Sestu:2015,Soufiani:2015,Ishihara:1994,Cooke:2015,YamadaAppPE:2014,Fang:2015,Huang:2013,Tahara:2016,Wolf:2014}.  A parameter that has received considerable attention recently is the exciton binding energy ($E_b$), which determines the primary mechanism of carrier transport during device operation.  Unlike inorganic solar cell materials such as III-V semiconductors, in which sharp exciton resonances greatly simplify extraction of $E_b$, the solution-processed organometal halide perovskites possess substantial broadening due to defects \cite{Yin:2014} and intrinsic dynamic disorder associated with the freedom of orientation of methylammonium ions \cite{Bakulin:2015,Ma:2015}.  The broadening associated with these effects, which has been found to vary from 30 meV at cryogenic temperatures to 60 meV at room temperature \cite{YamadaIEEE:2015}, is comparable to (or larger than) the exciton binding energy.  This broadening complicates the extraction of $E_b$ from absorption or ellipsometry techniques \cite{Sestu:2015}.  In addition, while photoluminescence provides an effective probe of defect-induced localization and recombination processes \cite{He:2016,Wehrenfennig:2014}, these techniques cannot isolate transitions relevant to absorption within an operational solar cell and can be hampered by incomplete phase transitions \cite{YamadaIEEE:2015,Wehrenfennig:2014}.  Extraction of $E_b$ from magneto absorption techniques is also complicated by the strong broadening in the organometal halide perovskites, resulting in the need to use large magnetic fields ($\geq$60 Tesla) \cite{Miyata:2015,Galkowski:2016}.  Due to these challenges, even for the most widely studied material CH$_3$NH$_3$PbI$_3$, the value of $E_b$ has been quite controversial, with reports ranging from 2~meV to 55~meV \cite{Wu:2014,Savenije:2014,Sun:2014,Tanaka:2003,Hirasawa:1994,Miyata:2015,Galkowski:2016,Lin:2014,SabaNatComm:2014,Even:2014,DInnocenzo:2014,YamadaIEEE:2015,Sestu:2015,Soufiani:2015,Ishihara:1994,Cooke:2015}. 

Here we apply femtosecond four-wave mixing spectroscopy (FWM) to probe excitonic resonances in CH$_3$NH$_3$PbI$_3$.  In this technique, two coherent ultrafast laser pulses described by electric fields $\vec{E_1}$ and $\vec{E_2}$ with wave vectors $\vec{k_1}$ and $\vec{k_2}$ excite a polarization component that emits light with wavevector 2$\vec{k_2}$-$\vec{k_1}$ (see Fig.~1).  Measurement of the emitted light versus the detection photon energy and the time delay between the two incident laser pulses provides a wealth of information about the optical transitions within the system under study, including the resonance frequencies and the time scale for scattering processes that cause decay of coherence (the so-called dephasing time, $T_2$)\cite{March:unpub}.  For semiconductors, FWM is highly sensitive to excitons because the signal at each pulse delay is integrated over the coherence lifetime of the transition. Excitons, being charge-neutral spatially-localized excitations, have much longer $T_2$ times than delocalized free carrier and band tail transitions \cite{ShahBook,Segschneider:1997}.  This sensitivity to excitonic effects has been exploited to observe the fundamental exciton in LT-GaAs,\cite{Webber:2014} in which the optical band edge is strongly broadened by As$_{\textrm{Ga}}$ antisite defects preventing the observation of any signature of the exciton in linear absorption, as well as the exciton resonance at the spin-orbit split-off band gap in InP \cite{Hall:2002}, which is masked in linear spectroscopy by strong degenerate interband transitions associated with the lower-energy band gaps.  

Our experiments reveal two discrete resonances below the band gap of the CH$_3$NH$_3$PbI$_3$ thin films, which we attribute to the free exciton transition and a defect-bound exciton transition.  The simultaneous observation of both free and localized excitons using this absorption-based nonlinear optical technique shows that both types of exciton can contribute to absorption during operation of a solar cell.  At 10 K, the excitonic resonances we observe occur at 1.629~eV and 1.613~eV. The 16~meV energy separation is attributed to the binding energy of excitons to point defects within the perovskite film \cite{Yin:2014}, and is in reasonable agreement with the recent measurement of an exciton localization energy of 17~meV in CH$_3$NH$_3$PbI$_{3-x}$Cl$_x$ using photoluminescence and THz techniques \cite{He:2016}.  Our measured transition energies correspond to an exciton binding energy of 13 meV (29 meV) for the free (defect-bound) exciton at 10~K, with a negligible temperature dependence up to 40~K.  The low-temperature free exciton binding energy we measure is in reasonable agreement with the value of 16~meV determined at 2~K from recent magneto-optical studies at high magnetic fields \cite{Miyata:2015,Galkowski:2016}.  Together with the difficulty in fully characterizing the dielectric constant including the potential contributions of phonons \cite{Even:2014,Soufiani:2015} and related dispersion \cite{Cooke:2015}, the close proximity of the free and bound exciton states we observe has likely contributed to the broad range of exciton binding energies reported in the literature \cite{Wu:2014,Savenije:2014,Sun:2014,Tanaka:2003,Hirasawa:1994,Miyata:2015,Galkowski:2016,Lin:2014,SabaNatComm:2014,Even:2014,DInnocenzo:2014,YamadaIEEE:2015,Sestu:2015,Soufiani:2015,Ishihara:1994,Cooke:2015,Huang:2013}.  

\begin{figure}[htb]\vspace{0pt}
    \includegraphics[width=17.0cm]{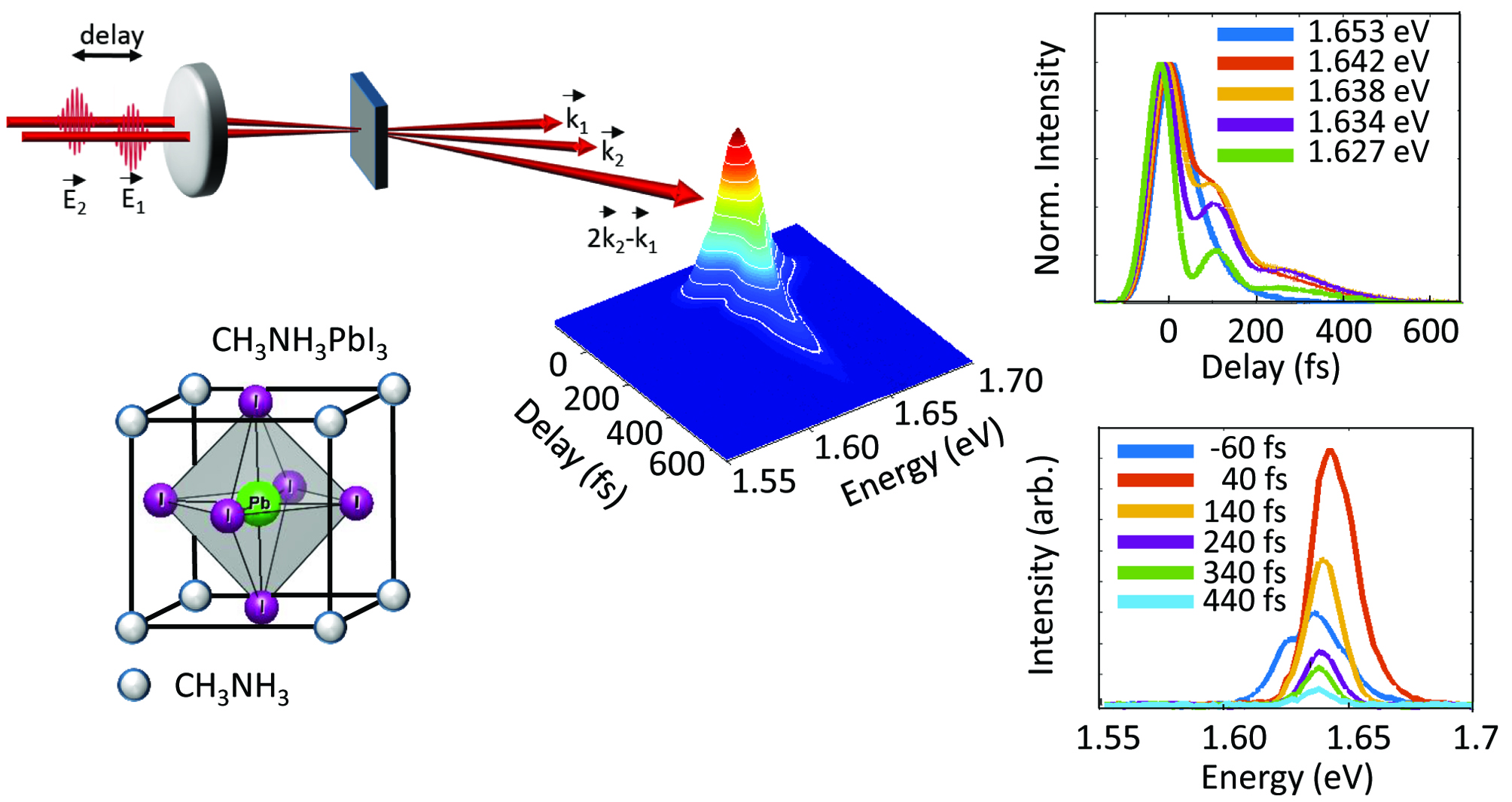}
    \caption{{\bf Upper left}, schematic of four-wave mixing spectroscopy:  Two 40~fs input pulses $\vec{E_1}$ and $\vec{E_2}$ excite a carrier density grating, and the self-diffracted signal along $2\vec{k_2}-\vec{k_1}$ is measured using a monochromator and photomultiplier detector as a function of the time delay.  {\bf Lower left}, methylammonium lead iodide perovskite ABX$_3$, with A= CH$_3$NH$_3$, B=Pb, X=I. {\bf Center} FWM response of the CH$_3$NH$_3$PbI$_3$ sample at 10~K for excitation at 1.653~eV, 11~meV above the band gap. {\bf Upper Right}, FWM signal at fixed values of detection photon energy, showing a smooth decay (oscillations) for energies above (below) band gap.  {\bf Lower right}, FWM spectrum at fixed values of time delay, indicating multiple spectrally-distinct signal contributions.}
    \label{fig:figure1}
\end{figure}

The four-wave mixing signal measured on the CH$_3$NH$_3$PbI$_3$ sample at 10~K is shown in Fig.~\ref{fig:figure1}.  For these results, the center photon energy of the laser spectrum was 1.653~eV, 11 meV above the band gap determined from linear absorption measurements (Fig.~S1).   The FWM spectrum exhibits a strong dependence on the time delay between the exciting laser pulses, as shown in the lower right panel of Fig.~\ref{fig:figure1}. For time delays on the rising edge of the FWM signal, the spectrum consists of three peaks.  The lowest energy peak has a maximum between 1.62~eV and 1.63~eV, with a tail on the low energy side.   The other two peaks occur at approximately 1.635~eV and 1.648~eV.  The FWM signal reaches its maximum intensity around zero delay, characterized by a broad response encompassing a range of energies above and below the band gap.  For longer delays, the signal narrows spectrally and decays on a time scale of a few hundred femtoseconds.  Cuts through the two-dimensional FWM results versus time delay are displayed in the upper right panel of Fig.~\ref{fig:figure1}. The signal exhibits a smooth temporal decay for detection energies above the band gap, while for energies below the band gap oscillations appear.  These oscillations resemble beating effects observed in FWM studies in the presence of simultaneous responses due to bound and free excitons in GaAs superlattices \cite{Koch:1993} and exciton and trion resonances in doped CdTe quantum wells \cite{Gillot:1999}. Together with the observation of three peaks in the four-wave mixing spectrum, this suggests that there are multiple, distinct contributions to the measured FWM response of the perovskite film.

\begin{figure}[htb]\vspace{0pt}
    \includegraphics[width=17cm]{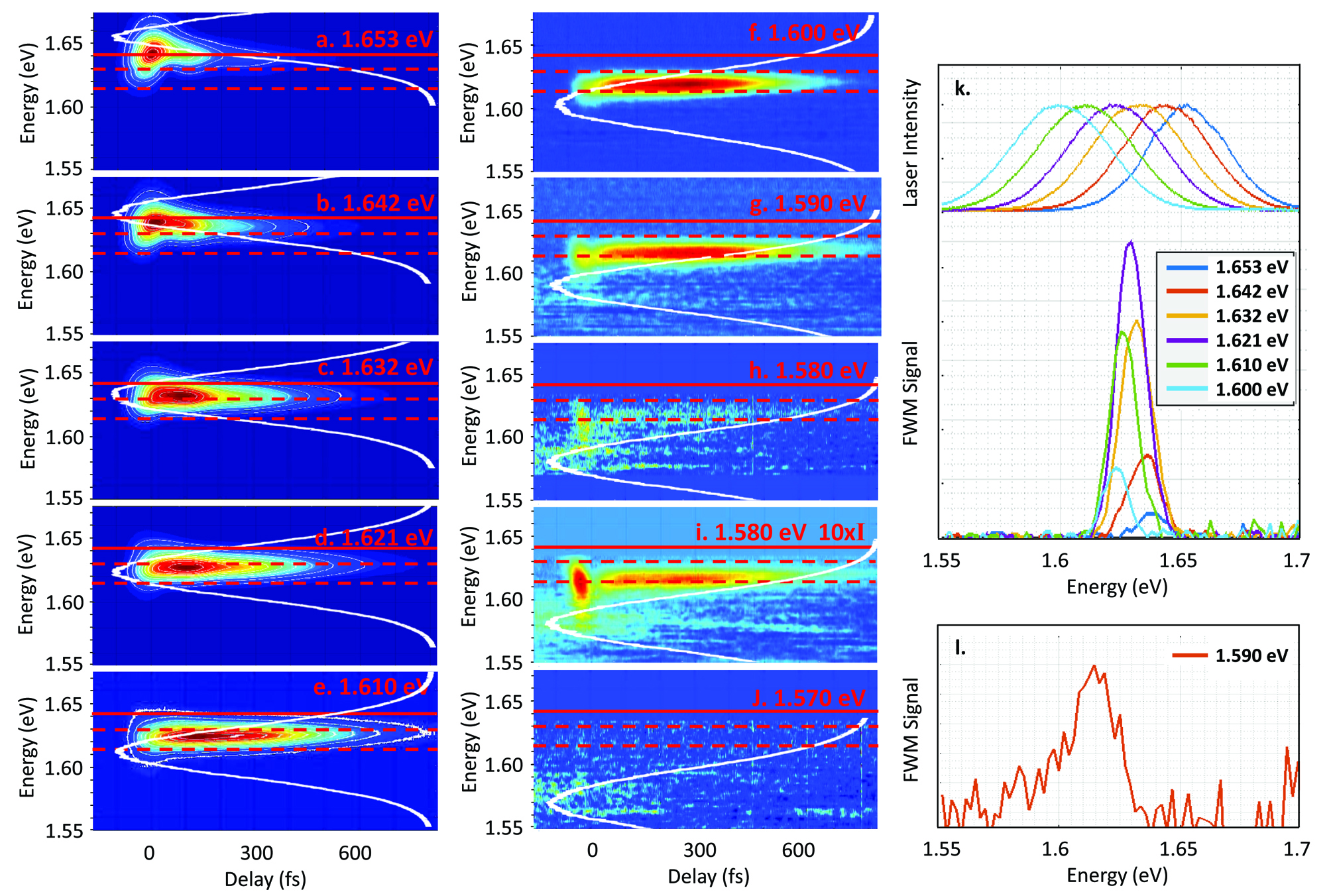}
    \caption{{\bf a-j} FWM response of CH$_3$NH$_3$PbI$_3$ for varying laser tuning relative to the band gap.  The FWM signal is displayed using a colored contour scale.  The white curve overlaid on the FWM results indicates the laser spectrum.  The solid red line indicates the 10~K band gap determined from linear absorption.  The two dashed lines indicate the resonance energies of 1.613~eV and 1.629~eV determined for the defect-bound and free exciton transition, respectively.  The defect-bound exciton appears as a weak, spectrally-isolated peak when the laser is tuned well below band gap ({\bf g-i}).   The free exciton resonance, which produces a much stronger FWM signal than the bound exciton, is evident from the dependence of the FWM signal amplitude on laser tuning ({\bf k}).  In {\bf k}, the laser spectra are vertically offset from the FWM spectra.  The free exciton dominates the four-wave mixing signal for time delays longer than 200~fs: results for 540~fs delay are shown in {\bf k}. {\bf l} FWM spectrum at zero delay for excitation at 1.590~eV, showing the defect-bound exciton peak.  The intensity of the 1.590~eV excitation laser pulses was 10 times larger for the results in {\bf i}, permitting the defect-bound exciton to be observed more clearly.}
    \label{fig:figure2}
\end{figure}

Further insight into the transitions contributing to the FWM response may be gained by performing measurements for which the laser pulses used to excite the sample are tuned relative to the band gap energy.  The results of these experiments are shown in Fig.~\ref{fig:figure2}.  In each panel, the laser spectrum is indicated by the white curve overlaid on the FWM signal, and the solid red line shows the band gap energy.   The existence of a separate contribution to the FWM signal below the band gap is already apparent at early delays from the shape of the contours in the results for excitation at 1.653~eV in Fig.~\ref{fig:figure2}a.  The broadband signal at higher energies in Fig.~\ref{fig:figure2}a is attributed to free carrier transitions \cite{Becker:1988} and transitions within the Urbach band tail \cite{Wolf:2014,YamadaAppPE:2014,Segschneider:1997}.  As the laser is tuned below the band gap energy (Fig.~\ref{fig:figure2}b-Fig.~\ref{fig:figure2}i), the peak below the band gap grows in intensity relative to the free carrier response.  When the laser is tuned to 1.580~eV, only the tail of the laser spectrum weakly excites the low-energy resonance, isolating it from other contributions, indicating a center energy of 1.613~eV and a spectral width of $\sim$18~meV (Fig.~\ref{fig:figure2}l).  When the laser pulses are tuned to 1.570~eV or lower energies, the laser spectrum has negligible overlap with the 1.613~eV peak, and no FWM signal is detected.   The discrete resonance we observe in the FWM signal at 1.613~eV is attributed to an excitonic transition within the CH$_3$NH$_3$PbI$_3$ film.

\begin{figure}[htb]\vspace{0pt}
    \includegraphics[width=14cm]{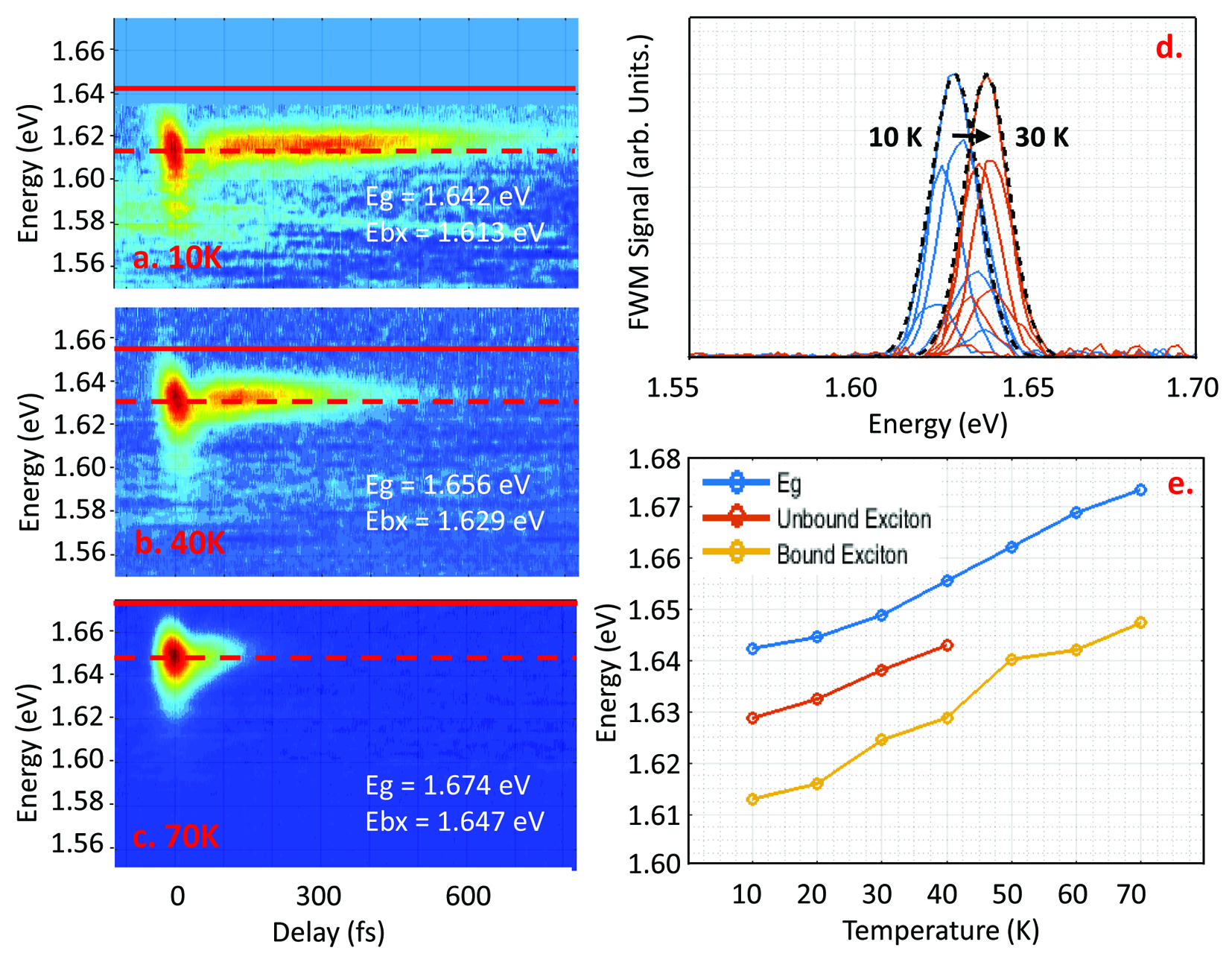}
    \caption{FWM response of the defect-bound exciton at ({\bf a}) 10~K, ({\bf b}) 40~K and ({\bf c}) 70~K , with the 
corresponding bandgap indicated by the solid line. {\bf d}  FWM spectrum at 540~fs delay for a variety of spectral tunings, showing the variation in the resonance energy of the free exciton as the temperature is increased from 10~K (blue curves) to 30~K (red curves).  The black dashed curves indicate a Gaussian fit to the resonance at each temperature. {\bf e}  The exciton transition energies deduced from the FWM results as a function of temperature.  Both exciton resonance energies follow the temperature dependence of the band gap within the experimental uncertainty, indicating no discernible temperature dependence of the exciton binding energies within the accessible range.}
    \label{fig:figure3}
\end{figure}

The variation of the magnitude of the FWM signal as the laser pulses are tuned indicates that the resonance feature at 1.613~eV is not the only discrete resonance below the band gap.  A stronger resonance occurs at higher energies, as shown in Fig.~\ref{fig:figure2}k, where the FWM spectrum at 540~fs delay is displayed color-coded with the corresponding laser spectrum in the upper panel.  This second resonance has a maximum at 1.629~eV, with a width of 15~meV.  The variation of the spectral content of the FWM signal as the laser is tuned indicates inhomogeneous broadening.  Both of the resonances at 1.613~eV and 1.629~eV have spectral widths smaller than the laser pulse bandwidth and have transition energies that are fixed as the center photon energy of the laser spectrum is varied.  These features are reflective of discrete, excitonic transitions within the semiconductor.  The strong 1.629~eV resonance is attributed to the free exciton transition, while the weaker resonance at 1.613~eV is attributed to an exciton bound to defects within the perovskite film.  From a comparison of the transition energy of each exciton with the band gap energy, we determine the binding energies of these excitons to be 13~meV and 29~meV, respectively.  The temperature-dependence of the FWM response from the two excitons is shown in Fig.~\ref{fig:figure3}.  The free exciton resonance energy could only be determined up to 40~K due to the limited tuning range of the laser relative to the band gap at higher temperatures.  The spectral position of the free and defect-bound exciton resonances both track the temperature dependence of the band gap energy, indicating no discernible dependence of the binding energies of the excitons on temperature below 40~K.

The binding energy of 13~meV we measured for the free exciton is in line with the value of 16 meV found from recent magneto-optical measurements at 2~K \cite{Miyata:2015,Galkowski:2016}.  In Ref.~\cite{Miyata:2015} and Ref.~\cite{Galkowski:2016}, large magnetic fields (up to 140 Tesla) enabled the observation of the 2s exciton transition, constraining the value of the binding energy over a much narrower range than in earlier magneto-optical experiments \cite{Tanaka:2003,Hirasawa:1994}.  In our FWM experiments, the free exciton was found to contribute much more strongly than the bound exciton.  Together with the higher sensitivity of FWM to excitonic resonances than linear absorption experiments, this accounts for the fact that the bound exciton was not observed in Refs.~\cite{Miyata:2015,Galkowski:2016}.  The binding energy of the free and defect-bound exciton resonances we observe are both within the large spread of reported values of 2-55~meV for CH$_3$NH$_3$PbI$_3$,\cite{Wu:2014,Savenije:2014,Sun:2014,Tanaka:2003,Hirasawa:1994,Miyata:2015,Galkowski:2016,Lin:2014,SabaNatComm:2014,Even:2014,DInnocenzo:2014,YamadaIEEE:2015,Sestu:2015,Soufiani:2015,Ishihara:1994,Cooke:2015} shedding some light on the controversy surrounding $E_b$ in recent years . The uncertainty in determining the dielectric constant, including both the choice of appropriate spectral range \cite{Huang:2013} and the complexity tied to the role of lattice vibrations \cite{Even:2014,Cooke:2015,Soufiani:2015}, has also contributed substantially to the wide range of reported binding energies since this value is essential to the modeling of absorption and photoluminescence.  Since the transition energies of the free and defect-bound excitons are determined directly using FWM spectroscopy, the dielectric constant is not required to determine the binding energies.  We note that in a recent work by Even \textit{et al.}, the dielectric constant was taken as a fitting variable in modeling the linear absorption spectrum, yielding a free exciton binding energy of 13~meV at 80~K, in agreement with the value reported here. 

The separation between the free and defect-bound exciton resonances is 16~meV.  We attribute this to the binding energy of excitons to point defects within the CH$_3$NH$_3$PbI$_3$ film.  The energy levels tied to a variety of point defects were calculated using density functional theory techniques by Yin \textit{et al.} \cite{Yin:2014}.  All point defects with low formation energies were found to be shallow defects, with energies less than 50~meV away from the band extrema.  The reported transition energies in Ref.~\cite{Yin:2014} suggest that the dominant donor MA$_i$ may be responsible for exciton localization at low temperatures in our sample.  The small 16~meV binding energy of these defects is consistent with a similar coherence decay time observed for the free and defect-bound excitons in our experiments (Fig.~S2), which is a signature of weak localization \cite{Siegner:1992}.  Our findings are also in line with an exciton localization energy of 17~meV determined from photoluminescence and THz studies on CH$_3$NH$_3$PbI$_{3-x}$Cl$_x$ \cite{He:2016}.   Static disorder associated with the frozen methylammonium cations at low temperatures \cite{Even:2014,Ma:2015} may contribute to inhomogeneous broadening of the exciton resonances, which is reflected by the measured linewidths being $\sim$5 times larger than the homogeneous width estimated from the duration of the coherent emission.  

Multiple emission peaks have also been observed in photoluminescence experiments on single crystals of CH$_3$NH$_3$PbI$_3$ \cite{Fang:2015}, however the energy separation between the emitting exciton resonances was considerably larger than that found using FWM techniques in this work.  As FWM detects the states participating in absorption, this suggests that carriers relax into deeper energy levels prior to radiative recombination.  Our observations, which verify the existence of weakly localized exciton states in this system, are in agreement with recent power-dependent photoluminescence studies which determined that excitons recombine radiatively from trap states \cite{He:2016} and transient photoconductivity experiments that indicated a resonant transition below the band gap tied to localized states \cite{Tahara:2016}.  Our findings are also consistent with earlier four-wave mixing experiments in disordered III-V superlattices revealing simultaneous signatures of free excitons and excitons bound to neutral Carbon acceptors \cite{Koch:1993}.     

The application of coherent nonlinear optical spectroscopy to CH$_3$NH$_3$PbI$_3$ reported here complements and extends the numerous studies of this system using linear optical techniques in recent years, for which the substantial broadening inherent to the disordered organic inorganic perovskite materials obscures fundamental photophysical properties such as excitonic effects, leading to challenges in assessing crucial parameters for solar cell design.  Since FWM preferentially detects optical species possessing long coherence decay times, this technique is highly sensitive to excitonic transitions, a feature that has been exploited for studying excitons in clean and disordered III-V and II-VI semiconductors over more than two decades \cite{ShahBook,Webber:2014,Hall:2002,Koch:1993,Gillot:1999}.  The direct observation of the resonance energies of both free and defect-bound excitons in CH$_3$NH$_3$PbI$_3$, eliminating the need to rely on assumed values of the dielectric constant for determination of the associated binding energies, aids in clarifying the broad range of values of the exciton binding energy reported in recent years as the relative role of free and defect-bound excitons will vary depending on whether absorption or photoluminescence based detection schemes are utilized.  The observation of defect-bound excitons using a technique based on nonlinear absorption indicates that both free and localized exciton species play a role in an operational solar cell.  Our findings demonstrate the power of FWM spectroscopy for studying excitonic properties and the influence of disorder in the organometal halide perovskite class of semiconductors, and will aid in improving solar cell device performance using this material system.


\section{Acknowledgements}     
This research is supported by the Natural Sciences and Engineering Research Council
of Canada. 

\section{Author contributions}
C.C. carried out the fabrication of the solution-processed CH$_3$NH$_3$PbI$_3$ film, under the guidance of I.G.H.  K.C.H. conceived and designed the FWM experiments. D.W. contributed to developing the FWM apparatus.  S.M. performed the FWM experiments with assistance by D.B.R. and analyzed the data.  S.M. and K.C.H. wrote the manuscript with input from all authors.

\section{Additional information}
Supplementary information is available in the online version of the paper.    Correspondence and requests for materials should be addressed to K.C.H.


\section{Competing financial interests}
The authors declare no competing financial interests.

\clearpage

\clearpage

\clearpage
\clearpage

\end{document}